\documentclass[a4paper]{article}
\usepackage{Odyssey2022}
\usepackage{epsfig,amssymb,amsmath}
\ninept

\usepackage{mathtools, amsfonts}
\usepackage{mathrsfs}
\usepackage{multicol}
\usepackage{multirow}
\usepackage{tikz}
\usepackage{tabu}
\usepackage{booktabs, tabularx}
\usepackage[referable]{threeparttablex}
\usepackage{stfloats}
\usepackage{comment}
\usepackage{xcolor}
\usepackage{hyperref}
\usepackage{caption} 
\usepackage{comment}
\usepackage{bm}
\usepackage{array}
\usepackage{hyperref,tablefootnote,footnotehyper}
\usepackage[inkscapearea=page]{svg}
\usepackage{adjustbox}
\usepackage{url} 
\usepackage{breakurl}

\newcommand{\norm}[1]{\left\lVert#1\right\rVert}

\def\x{{\mathbf x}}
\def\z{{\mathbf z}}

\renewcommand{\vec}[1]{\boldsymbol{\mathrm{#1}}}
\newcommand{\mtx}[1]{\boldsymbol{\mathrm{#1}}}
\newcommand{\transp}{\ensuremath{^\mathsf{T}}}

\makeatletter
\newcommand{\thickhline}{%
    \noalign {\ifnum 0=`}\fi \hrule height 1pt
    \futurelet \reserved@a \@xhline
}
\newcolumntype{"}{@{\hskip\tabcolsep\vrule width 1pt\hskip\tabcolsep}}
\makeatother

\title{Magnitude-aware Probabilistic Speaker Embeddings} 

\name{{\em \textit{Nikita Kuzmin$^{1*}$}\thanks{${}^{*}$Equal contribution.}, \textit{Igor Fedorov$^{2*}$}\footnotemark[1], \textit{Alexey Sholokhov$^3$}}}

\address{$^1$Lomonosov Moscow State University, Moscow, Russia \\
$^2$National Research University ``Higher School of Economics'', Moscow, Russia\\
$^3$Independent researcher \\
\small \tt s02170136@gse.cs.msu.ru, idfedorov@edu.hse.ru, sholokhovalexey@gmail.com \\
}

\begin{document}
\maketitle

\setlength{\abovedisplayskip}{3pt}
\setlength{\belowdisplayskip}{3pt}

\begin{abstract}
Recently, hyperspherical embeddings have established themselves as a dominant technique for face and voice recognition. Specifically, Euclidean space vector embeddings are learned to encode person-specific information in their direction while ignoring the magnitude. However, recent studies have shown that the magnitudes of the embeddings extracted by deep neural networks may indicate the quality of the corresponding inputs. This paper explores the properties of the magnitudes of the embeddings related to quality assessment and out-of-distribution detection. 
We propose a new probabilistic speaker embedding extractor using the information encoded in the embedding magnitude and leverage it in the speaker verification pipeline. 
We also propose several quality-aware diarization methods and incorporate the magnitudes in those. Our results indicate significant improvements over magnitude-agnostic baselines both in speaker verification and diarization tasks. 
\end{abstract}


\section{Introduction}

In recent years, deep learning-based methods have shown impressive performance in speaker verification and diarization. However, recognition in the wild is still difficult due to the large variability in speech signals acquired in unconstrained conditions. This variability is caused by different channels or recording environments involving background noise and reverberation. 

Therefore, the audio segments provided as an input to a speaker recognition system are not always appropriate for recognition. A typical speaker recognition system operates under the assumption that the input audio contains recognizable information, and this assumption usually holds during training because the training datasets are specifically collected and filtered in this way. However, this is not always the case after deployment, as the input signal may be of low quality or may not even contain human speech, which could lead to incorrect predictions because the embedding extractor has never seen such inputs before.

One solution to this issue is to train an extra model to assess the quality of the input recordings. Thus, ground-truth labels must be provided for a training dataset, including recordings of varying quality. However, the quality labels provided by humans or those derived from the signal-to-noise ratio (SNR) or signal duration \cite{Mandasari2015-qmf, Villalba2013-bn, Lavrentyeva2020-quality} may not be the most relevant characteristics in terms of recognition performance.

\begin{figure}[t!]
 \centering
 \adjustbox{trim=0 3mm 0 0}{
 \includegraphics[scale=0.35]{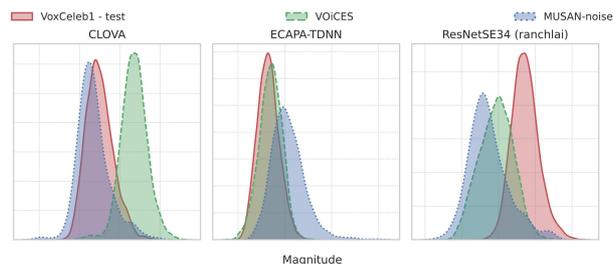}
 }
 \caption[cykablyat]{Distributions of embeddings magnitudes for three publicly available pre-trained embedding extractors. Left\footnotemark[1]: ResNetSE34 by CLOVA \cite{Heo2020-clova}.
 Center\footnotemark[2]: ECAPA-TDNN \cite{Desplanques2020-ecapa} from the SpeechBrain package \cite{speechbrain}.
 Right\footnotemark[3]: ResNetSE34 by \cite{ranchlai}. Only in the last case the magnitudes serve as weak descriptors of speech quality. Best viewed in color.}
 \label{fig:clova-ecapa}
\end{figure}
\footnotetext[1]{\scriptsize \url{https://github.com/clovaai/voxceleb_trainer}}
\footnotetext[2]{\scriptsize \url{https://huggingface.co/speechbrain/spkrec-ecapa-voxceleb}}
\footnotetext[3]{\scriptsize \url{https://github.com/ranchlai/speaker-verification}}

Another group of related studies revealed how to obtain the information about the input quality directly from the extracted embeddings. In \cite{Ranjan2017-l2} it was observed that face embeddings learned using the softmax loss tend to have a smaller magnitude \footnotemark[4] \footnotetext[4]{``Magnitude'' hereafter stands for $L_2$-norm.} for the inputs of lower quality (\emph{e.g.} blurry images). Further, \cite{meng2021magface, Scott2021-vmf} proposed new classification-based losses to explicitly encourage magnitudes to represent feature confidence in the embedding space. An appealing property of magnitudes is providing useful information for the predictions confidence estimation without any extra cost. 

There also exists a concept of probabilistic embeddings, which naturally provides the quality estimates as variance-related probability distribution parameters. Such embeddings are well studied in both face recognition \cite{shi2019PFE, Chen2021FastAR, li2021spherical} and speaker recognition research \cite{Brummer2018-gme, Silnova2020-meta}.

The main contributions of this paper are as follows.
\begin{itemize}
  \item First, inspired by \cite{meng2021magface}, we train speaker embedding extractor with interpretable embedding magnitudes and examine properties of the resulting embeddings pertaining to the speech recording quality.
  \item We introduce a specific form of probabilistic embeddings with Gaussian precision interpretation for magnitudes and evaluate it for the speaker verification task. 
  
  \item We also explore the ability of embeddings magnitudes to filter out non-speech or low-quality speech segments. Based on this, we propose a two-step speaker diarization method that relies on a subset of reliable segments to robustly identify initial clusters. 
  
  \item Last, we modify a popular diarization method known as VBx \cite{landini2020bayesian} by enabling uncertainty propagation in this model.
\end{itemize}

\section{Background}


\subsection{Magnitude-aware embeddings} 

\label{sec:Norm-Aware approaches}


Earlier works in face recognition identified an interesting property of embeddings trained with the softmax loss function \cite{Ranjan2017-l2, Parde2016-conv}. 
It was found that embedding's magnitude can be a predictor for image quality, where the quality refers to various factors affecting the ability to identify a person such, as a blur, low resolution, extreme pose, or bad lighting.
Similar observations were later made in \cite{Scott2021-vmf}. As discussed in \cite{Scott2021-vmf}, 
the explanation could be that since the magnitude affects the peakedness of the class posterior distribution, hard and noisy samples will be forced to have smaller magnitudes during training. 
Moreover, these authors also found that this effect exists for angular losses, such as ArcFace \cite{deng2018arcface}, despite their magnitude-agnostic nature. 




Based on these findings, we studied magnitudes of a few modern top-performing publicly available speaker embedding extractors, all of which were trained with the softmax-based losses. For this estimation we employed the VoxCeleb1 test set \cite{voxceleb1}, the evaluation set from the VOiCES from a Distance Challenge 2019 \cite{voices} and the MUSAN noise collection \cite{musan2015}. These datasets represent clean speech, noisy speech, and non-speech recordings, respectively. 
Figure \ref{fig:clova-ecapa} shows the distributions of magnitudes for these datasets. Surprisingly, only one of the embedding extractors allows interpretation of magnitudes as quality estimates. This leads to the conclusion that such interpretation is not provided ``by default''.

It is worth mentioning that the magnitudes of speaker embeddings were already successfully applied for the voice activity detection task \cite{kwon2021look}. However, this work lacks any qualitative analysis of embedding magnitudes properties. To fill this gap, we present such analysis in section \ref{sec:analysis}. 

Based on these observations, we decided to train the embedding extractor using the recently proposed MagFace loss \cite{meng2021magface} designed to explicitly learn quality-aware representations. As a reference, we choose the ArcFace (also referred to as AAM-Softmax \cite{chung2020in}) as a loss without an explicit mechanism to affect the embedding magnitude. 





ArcFace \cite{deng2018arcface}, is among the most widely adopted loss functions in speaker recognition. This loss function can be obtained from the softmax loss by normalizing the classifier weights and input embeddings and by introducing the \emph{constant} angular margin into the target logits. Although this loss function is magnitude-agnostic, it still features the confidence propagation behavior\footnote{But not necessary, \emph{e.g.}, the pre-trained ECAPA-TDNN \cite{Desplanques2020-ecapa} from SpeechBrain \cite{speechbrain} lacks this property, despite having AAM-Softmax as a training objective.}. 

MagFace \cite{meng2021magface} loss function can be seen as a generalization of ArcFace with an \emph{adaptive} angular margin depending on the magnitude of the input embedding $\vec{x}$. The loss can be formulated as follows:
\begin{equation} \label{eq:loss}
    \mathcal{L}_{Mag} = - \frac{1}{N} \sum_{i=1}^N\log \frac{e^{\psi(\theta_{y_i}^i)}}{e^{\psi(\theta_{y_i}^i)} + \sum\limits_{j \neq y_i} e^{s \cos \theta_{j}^i}} + \lambda_g g(||\vec{x}_i||),
\end{equation}
where $i$ runs over the training batch of size $N$, and $\theta_{j}^i \in \mathbb{R}$ represents the angle between the $j$-th classifier weight and $d$-dimensional embedding vector $\x_i \in \mathbb{R}^d$ with the corresponding label $y_i$. In addition, $\psi$ denotes the angle function depending on the base loss, which is defined as $\psi(\theta_{y_i}^i) = s\cos(\theta_{y_i}^i + m)$ for ArcFace, where $m$ is a margin and $s$ is a fixed scale factor. In MagFace the margin $m \equiv m(\norm{\x_i})$ is a strictly increasing convex function of the embedding's magnitude. Additionally, the MagFace loss has a second term, weighted by $\lambda_g$, where $g$ is a strictly decreasing convex function. 


\subsection{Probabilistic embeddings} 

\label{subsec:probemb}
In the representation learning, the extracted vector representations are conventionally called \emph{embeddings}. These embeddings usually are point-wise, or \emph{deterministic}, estimates of some latent variable in the representation space. Alternatively, one can consider \emph{probabilistic} representations, usually in a form of probability distributions in the latent space \cite{karpukhin2022-probabilistic}. Such representations are of sufficient interest, as they can capture more information than simple point-wise estimation, \emph{e.g.} quantify the data uncertainty for input sample, and provide more flexibility to define similarity measures. 



Recently, probabilistic embeddings gained popularity for face recognition in unconstrained conditions where the recognition systems may heavily suffer from low-quality inputs \cite{shi2019PFE, Chen2021FastAR, li2021spherical, Chang2020-DUL}.
For instance, the authors of \cite{shi2019PFE} proposed to represent the input images by Gaussian distributions. These distributions are parameterized by diagonal covariance matrices, estimated by an additional branch to the embedding extractor. In this architecture, the covariance matrix represents the dimension-wise uncertainty of face embeddings treated as hidden variables. 




In this work, we consider a particular instance of probabilistic embeddings known as  \emph{meta-embeddings} \cite{Brummer2018-gme} that provide the likelihood distribution for embeddings treated as hidden variables.
While the choice of particular functional form of meta-embeddings is flexible, \cite{Brummer2018-gme} proposed \emph{Gaussian meta-embeddings} (GME) as a tractable solution convenient for practical application.
Later, similar to \cite{shi2019PFE}, the authors of \cite{Silnova2020-meta} proposed an architecture to extract meta-embeddings by introducing an additional branch whose outputs are interpreted as precisions for the Gaussian distribution. This model yielded a noticeable performance gains for the speaker diarization task. 



\section{Magnitude-aware Gaussian meta-embeddings}


In this work, we propose a probabilistic embedding extractor that takes an advantage of the information encoded in the magnitudes of deterministic embeddings.
Our approach relies on the assumption that embeddings magnitudes can serve as indicators of the input quality.
Specifically, we use the magnitude to define a precision parameter in the Gaussian meta-embeddings.


\subsection{From magnitude to precision} 
\label{subsec:norm2prec}

Our approach can be seen as a specific recipe to design the probabilistic embeddings in the form described in \cite{Brummer2018-gme} or \cite{Silnova2020-meta}. In this section, we illustrate the proposed idea for the Gaussian meta-embeddings introduced in \cite{Brummer2018-gme} while it is also applicable to the model \cite{Silnova2020-meta} which uses probabilistic linear discriminant analysis (PLDA) as the embedding prior.


\emph{Meta-embedding} $f(\z)$ is a likelihood function for the latent speaker identity variable $\z \in \mathbb{R}^d$, \emph{i.e.}:
\begin{equation} \nonumber
    f(\z) \propto p(x|\z), 
\end{equation}
where $x$ denotes the input audio recording represented by waveform, Mel-spectrogram, or vector embedding. 
Given the embedding prior $p(\z)$, one can compare possible inputs partitions by computing the corresponding likelihood ratios expressed in terms of meta-embeddings.
In practice, however, one needs to restrict the functional form of meta-embeddings for the sake of tractability. One convenient option is Gaussian meta-embeddings (GME) where $f$ has the form of the Gaussian likelihood function.

Our proposal can be summarized as follows: we resort to the isotropic precision matrix and use the embedding magnitude as its scale parameter. This leads to the following formulation of GME: 
\begin{equation} \nonumber
    f(\z;x) = \exp \left[-\frac{||\bm{\mu}||}{2}\z\transp\z + \z\transp \bm{\mu} \right],
\end{equation}
where $\bm{\mu}$ denotes the $d$-dimensional speaker embedding extracted from the speech recording $x$ by a deep neural network: $\vec{\mu} = \mathrm{net}(x)$. By using the decomposition $\bm{\mu} = \overline{\bm{\mu}} \cdot ||\bm{\mu}|| $ where $\overline{\bm{\mu}} $ denotes the length-normalized emebdding, one can see that $||\bm{\mu}||$ can be interpreted as a scalar precision parameter in the isotropic Gaussian.

The potential advantage of this approach is that it does \emph{not} require training a separate model, allowing to obtain the precision parameter without extra cost. This makes it different from similar models \cite{Silnova2020-meta, Garcia-Romero2020}, which augment the embedding extractor with an additional branch that provides an additional output. 

Given a pair of inputs $\{x_1, x_2\}$ representing the verification trial and assuming the standard Gaussian prior $p(\z) = \mathcal{N}(\vec{0}, \mtx{I})$, as in \cite{Brummer2018-gme}, this results in the following expression for the log-likelihood ratio (LLR) score:
\begin{multline} \nonumber
    \text{LLR}(x_1, x_2) = \frac{1}{2} \frac{(\vec{\mu}_1 + \vec{\mu}_2)\transp (\vec{\mu}_1 + \vec{\mu}_2)}{r_1 + r_2 + 1}  - \frac{1}{2} \frac{\vec{\mu}_1 \transp \vec{\mu}_1}{r_1 + 1} \\ - \frac{1}{2} \frac{\vec{\mu}_2 \transp \vec{\mu}_2}{r_2 + 1}  + \frac{d}{2}\log\frac{(r_1 + 1)(r_2 + 1)}{(r_1 + r_2 + 1)},
\end{multline}
where $r = ||{\bm{\mu}}||$ denotes the $L_2$-norm.
One may notice that this LLR is a shifted and scaled cosine similarity with both shift and scale parameters depending only on the embeddings magnitudes.


To ensure that the embedding magnitudes are indeed positively correlated with speech quality, \emph{e.g.} signal-to-noise ratio, we employ the MagFace loss \cite{meng2021magface} for training the extractor of deterministic embeddings. 
However, as we observed in the experiments, magnitudes may depend on the duration of speech recording. That is why we introduced a learnable parametric transform conditioned on the input duration. See also section \ref{sec:analysis} for related discussion.

\subsection{Compensation for utterance duration variability}
\label{subsec:transform}
We start with several assumptions and observations:
\begin{itemize}
    \item The embedding magnitude should be a good descriptor of recording quality allowing for Gaussian precision interpretation. We assume that there exists a \emph{monotonic mapping} from magnitudes to precisions.
    \item The qualitative analysis (see section \ref{sec:analysis}) indicates the \emph{descent} of embedding magnitudes with the increase of recording duration. This effect contradicts the confidence-based interpretation of magnitudes. We compensate for this by introducing the duration-based additive adjustment for the magnitude. 
    \item Finally, the adjusted magnitude should be transformed to the appropriate range for serving as the precision parameter.
\end{itemize}
Based on the above, we have chosen the following transformation rule:
\begin{equation}
    r_i = s \cdot (||\bm{\mu}_i|| + \gamma \cdot \min\{20, \text{len}(x_i)\}),
    \label{eq:norm-transform}
\end{equation}
where $s > 0$ indicates the global scaling parameter, which provides the appropriate range, $\text{len}(\cdot)$ stands for the recording duration (in seconds; we empirically constrain the maximum allowed duration to 20 seconds), and $\gamma > 0$ determines the impact of duration adjustment. It is important to note that described transformation is only needed for LLR scoring, while the raw magnitudes can be used for quality assessment purposes. The parameters in \eqref{eq:norm-transform} can be estimated from a development set of recordings.


\section{Speaker verification}
We start by evaluating the proposed models on the speaker verification task. 

\begin{table}[]
\scriptsize
\resizebox{\columnwidth}{!}{\begin{tabular}{ccccc}
\hline
\rule{0pt}{2.5ex} {ResNet-ArcFace}                                                                 & \multicolumn{2}{c}{ResNet-MagFace} & \multicolumn{2}{|c}{Large ResNet-MagFace}                                                \\ \hline
\rule{0pt}{2.5ex} {13.84M / 13.82 GFLOPs} &
 \multicolumn{2}{c}{{13.84M / 13.82 GFLOPs}} & \multicolumn{2}{|c}{{50.45M / 34.66 GFLOPs}}                                                \\ \hline
 \multicolumn{5}{c}{\rule{0pt}{2.5ex}c=7x7, f=128, s=1, p=3}  \\ \hline
\begin{tabular}[c]{@{}c@{}}\rule{0pt}{2.5ex}c=3x3, f=128, s=1, p=0\\ c=3x3, f=128, s=1, p=0\end{tabular} & $\times3$         & \multicolumn{1}{c|}{}         & \begin{tabular}[c]{@{}c@{}}\rule{0pt}{2.5ex}c=3x3, f=128, s=1, p=0\\ c=3x3, f=128, s=1, p=0\end{tabular} & $\times3$ \\ \hline
\begin{tabular}[c]{@{}c@{}}\rule{0pt}{2.5ex}c=3x3, f=128, s=2, p=0\\ c=3x3, f=128, s=2, p=0\end{tabular} & $\times4$          & \multicolumn{1}{c|}{}         & \begin{tabular}[c]{@{}c@{}}\rule{0pt}{2.5ex}c=3x3, f=128, s=2, p=0\\ c=3x3, f=128, s=2, p=0\end{tabular} & $\times8$  \\ \hline
\begin{tabular}[c]{@{}c@{}}\rule{0pt}{2.5ex}c=3x3, f=256, s=2, p=0\\ c=3x3, f=256, s=2, p=0\end{tabular} & $\times6$          & \multicolumn{1}{c|}{}         & \begin{tabular}[c]{@{}c@{}}\rule{0pt}{2.5ex}c=3x3, f=256, s=2, p=0\\ c=3x3, f=256, s=2, p=0\end{tabular} & $\times36$ \\ \hline
\begin{tabular}[c]{@{}c@{}}\rule{0pt}{2.5ex}c=3x3, f=256, s=2, p=0\\ c=3x3, f=256, s=2, p=0\end{tabular} & $\times3$         & \multicolumn{1}{c|}{}         & \begin{tabular}[c]{@{}c@{}}\rule{0pt}{2.5ex}c=3x3, f=256, s=2, p=0\\ c=3x3, f=256, s=2, p=0\end{tabular} & $\times3$ \\ \hline
\multicolumn{5}{c}{\rule{0pt}{2.5ex}Statistic Pooling \cite{snyder17-interspeech}} 
\\ \hline
\multicolumn{5}{c}{\rule{0pt}{2.5ex}FC (d=256)}                                                                                                                                                                                                                \\ \hline

\multicolumn{1}{c|}{\rule{0pt}{2.5ex}ArcFace Loss}                                                                                                     & \multicolumn{4}{c}{MagFace Loss}                                                              \\ \hline
\end{tabular}}
\caption{The description of architectures, where \textit{c} denotes a size of each convolutional filter, \textit{f} corresponds to a number of convolutional filters, \textit{s} denotes a stride size, \textit{p} denotes a padding size, and \textit{d} indicates the output embedding dimension. The Second line in the table shows models' efficiency characteristics. The first number in each pair denotes a number of parameters (in millions), the second one indicates amount of GFLOPs to process one second of audio.}
\label{tab:resnet_architectures}
\end{table}

\subsection{Implementation details}
\subsubsection{Architecture}
We implemented three different embedding extractors for comparison:
\begin{itemize}
    \item ResNet-ArcFace: ResNet trained with the ArcFace loss,
    \item ResNet-MagFace: ResNet trained with the MagFace loss,
    \item Large ResNet-MagFace: ResNet with more layers trained with the MagFace loss.
\end{itemize}
Architectures of these networks are shown in Table \ref{tab:resnet_architectures}.

As features, we used 80-dimensional log Mel-filterbank energies extracted with a window length of 25 ms and a frame-shift of 10 ms.

\subsubsection{Training}
All the speaker embedding extractors were trained on the development part of the VoxCeleb2 dataset \cite{Chung2018-vox2} using segments of 3 seconds randomly cropped from the original audios. The parameters of the embedding extractors were updated via the Ranger optimizer \cite{Ranger} with a cosine annealing learning rate scheduler. The minimum learning rate was set to $10^{-5}$ with a scheduler’s period equal to 100K iterations and the initial learning rate was equal to $10^{-3}$. The batch size was set to 64.

We applied an online augmentation strategy based on noises and room impulse responses from the MUSAN corpus \cite{musan2015}, following the recipe from \cite{Zeinali2019-BUT}. All the embedding extractors were trained for 600K iterations, and the checkpoint with the best performance on the test part of the VoxCeleb2 dataset was used for the evaluation.

Following \cite{meng2021magface}, we used the same functional form for the functions $m$ and $g$ in MagFace loss, Eq. (\ref{eq:loss}):
$$
g(a) = \frac{a}{n_u^2} + \frac{1}{a} \text{ and } m(a) = \frac{m_u - m_l}{n_u - n_l} (a - n_l) + m_l,
$$
where $n_u=110, n_l = 10, m_u = 1, m_l = 0.1$ are the maximum and minimum values of the magnitudes and margins, respectively, used for clamping during training.

The parameters of the precision transformation \eqref{eq:norm-transform} were estimated on the VoxCeleb2 test set by minimizing equal error rate on a custom verification protocol. Precisely, we firstly estimated the relative impact of duration based on desired correlation of duration-adjusted confidence with duration, and then selected the global scale to obtain the desired range of confidence values.

\begin{table*}[t]
\resizebox{\textwidth}{!}{%
\begin{tabular}{c|l|c|cc|cc|cc|cc}
\hline
\multirow{2}{*}{\textbf{Embedder}} &
  \multicolumn{1}{c|}{\multirow{2}{*}{\textbf{Loss}}} &
  \multirow{2}{*}{\textbf{Score}} &
  \multicolumn{2}{c|}{\textbf{VC1-test (cleaned)}} &
  \multicolumn{2}{c|}{\textbf{VC1-test (cleaned, 2s-pool)}} &
  \multicolumn{2}{c|}{\textbf{VB-pool}} &
  \multicolumn{2}{c}{\textbf{VOiCES Eval}} \\ \cline{4-11} 
                                                   & \multicolumn{1}{c|}{}    &                 & EER  & minDCF & EER  & minDCF & EER  & minDCF & EER  & minDCF \\ \hline
\multirow{4}{*}{ResNet} &
  \multirow{2}{*}{ArcFace} &
  cosine &
  1.13 &
  0.123 &
  3.47 &
  0.317 &
  6.42 &
  0.472 &
  4.50 &
  0.295 \\
 &
   &
  GME-LLR &
  \textbf{0.98} &
  \textbf{0.112} &
  \textbf{3.02} &
  \textbf{0.284} &
  \textbf{5.76} &
  \textbf{0.468} &
  \textbf{3.97} &
  \textbf{0.288} \\ \cline{2-11} 
 &
  \multirow{4}{*}{MagFace} &
  cosine &
  1.07 &
  0.109 &
  3.27 &
  0.292 &
  6.34 &
  0.462 &
  4.18 &
  0.263 \\
 &
   &
  GME-LLR &
  \textbf{0.92} &
  \textbf{0.105} &
  \textbf{2.85} &
  \textbf{0.277} &
  \textbf{5.40} &
  \textbf{0.452} &
  \textbf{3.61} &
  \textbf{0.256} \\ \cline{1-1} \cline{3-11} 
\multicolumn{1}{l|}{\multirow{2}{*}{Large ResNet}} &
   &
  cosine &
  0.66 &
  0.064 &
  2.59 &
  0.250 &
  6.09 &
  0.433 &
  4.14 &
  0.246 \\
\multicolumn{1}{l|}{} &
   &
  GME-LLR &
  \textbf{0.57} &
  0.064 &
  \textbf{2.30} &
  \textbf{0.237} &
  \textbf{5.93} &
  0.437 &
  \textbf{3.55} &
  \textbf{0.232} \\ \hline
\end{tabular}%
}
\caption{Results of speaker verification with three different embedding extractors. Performance is measured in terms of EER and minDCF with $P_{\mathrm{target}}=0.01$.}
\label{tab:verification}
\end{table*}

\subsection{Verification performance}
\subsubsection{Datasets}
\label{subsubsec:datasets}
We used two commonly adopted evaluation protocols: the original VoxCeleb1 test (cleaned) \cite{voxceleb1} and the evaluation set from the VOiCES from a Distance Challenge 2019 \cite{voices}. These tests represent the ``clean'' and ``noisy'' conditions, respectively. For the VoxCeleb1 test we created two additional copies with (a) both recordings cropped to 2 seconds from the beginning and (b) only the enrollment recording cropped to 2 seconds. All the three copies (original, (a) and (b)) were further concatenated to create a new protocol called \textit{2s-pool} with the purpose to demonstrate robustness to duration variability.

We also created a custom evaluation protocol using a parallel database of clean and noisy speech recordings called \mbox{VoiceBank} \cite{voicebank}. For each present identity, we took 20 random audios and created a Cartesian product on all taken audios, then converted this list into clean-vs-clean, clean-vs-noisy, and noisy-vs-noisy versions. We then concatenated them and eliminated duplicates. This pipeline produced $\sim 5.9$M pairs with $1.1\%$ positives. This protocol is referred to as \textit{VB-pool} and includes ``mixed'' conditions, intended to represent real-world applications with uncontrolled conditions.

\subsubsection{Metrics} 
We evaluated the speaker verification performance using two metrics: the \emph{equal error rate} (EER) and the \emph{minimum normalized detection cost function} (minDCF) with $P_{\mathrm{target}}=0.01$ \cite{Przybocki2004-nist}. 

\subsubsection{Results}
Table \ref{tab:verification} provides the speaker verification results. One can see that the GME back-end achieves lower EER on all the test sets. In addition, the MagFace objective provides slight improvements over ArcFace, on both clean and noisy data. Effects of moving from deterministic to probabilistic embeddings are precisely noticeable for ResNet-ArcFace and ResNet-MagFace networks, although, for Large ResNet-MagFace model the improvement is relatively smaller, especially in terms of minDCF, which even degrades for \textit{VB-pool}.



\section{Speaker diarization}





We now turn our attention to the speaker diarization task. This section presents how to improve the speaker diarization performance by using the \emph{segment-specific} quality scores. In particular, we used embedding magnitudes to provide such scores. The following two sections detalize our proposals.

\subsection{Quality-aware two-step pipeline}

First, we propose the following \emph{two-step} diarization method which may serve as a general template independent of a clustering back-end. 
The first step filters out the less reliable segments and finds initial clusters within the selected segments.
The second step uses all (or a subset of) available segments and refines their labels based on the clusters estimated in the first step. 

The proposed two-step method sorts embeddings magnitudes and divides embeddings into two subsets based on the percentile tuned on the development set. Then, the first step applies a general clustering algorithm with an unknown number of clusters to the set formed from embeddings with larger magnitudes. Intuitively, this should provide a better estimate for the number of clusters and their locations than one-step algorithms. The resulting labels were used as input to the second step of the algorithm, which has three possible implementations.

\textbf{Algorithm 2.1.} We computed the average embeddings (centroids) within each cluster found in the first step and assigned the remaining embeddings to the closest centroid using the similarity function.

\textbf{Algorithm 2.2.}  We applied a clustering algorithm with a fixed number of clusters (estimated in the first step) to \textit{all} embeddings.

\textbf{Algorithm 2.3.} 
We applied a clustering algorithm with a fixed number of clusters to the \textit{remaining} embeddings.

To solve the label ambiguity problem in (2.3) algorithm, we computed the centroids of the clusters from the first step and assigned the remaining embeddings to the closest centroid by cosine similarity. Finally, we found the dominating labels within each cluster found in the second step and assigned this label to all the embeddings within this cluster.

\subsection{Bayesian HMM with uncertainty propagation}

Second, we implement uncertainty propagation for the Bayesian hidden Markov model (HMM) back-end \cite{landini2020bayesian} in a similar way as described in \cite{Kenny2013-up, Lin2016-plda-up}.

We assume that the extracted utterance embeddings were generated by the PLDA model with the full-rank speaker subspace in the form of a two-covariance model \cite{Brummer2010-two-cov}. 
The model is specified by the following probability distributions:
\begin{equation}
    \begin{aligned}
        p(\vec{y}) & = \mathcal{N}(\vec{y}|\vec{0},\mtx{B}) \\ \label{eq:2cov-likelihood}
        p(\vec{x}|\vec{y}) & = \mathcal{N}(\vec{x}|\vec{y}, \mtx{W})
    \end{aligned}
\end{equation}
where the two matrices $\mtx{B}$ and $\mtx{W}$ model between- and within-speaker covariances. For each speaker, represented by latent variable $\vec{y}$, the distribution of embedding vectors $\vec{x}$ is modeled by a single Gaussian. These Gaussians are parameterized by speaker-dependent means $\vec{y}$ and a shared covariance matrix $\mtx{W}$. 

For clustering purposes, we construct a HMM where states correspond to speakers and the speaker-specific distributions are derived from the PLDA model \eqref{eq:2cov-likelihood}.
Assuming that $K$ speakers are present in the collection of embeddings, $\mtx{X}=\{\vec{x}_{1}, \dots, \vec{x}_{N}\}$, the model has $K$ latent speaker mean vectors $\mtx{Y} = \{\vec{y}_{1}, \dots, \vec{y}_{K}\}$, one for each speaker-specific Gaussian. 
This model can be seen as a particular case of the model in \cite{landini2020bayesian} that also supports the PLDA model with low-rank speaker subspace.

The uncertainty propagation mechanism is implemented by introducing an \emph{input-dependent} covariance matrix $\mtx{\Sigma}$:
\begin{equation}
    \begin{aligned}
        p(\vec{x}_i|\vec{y}) & = \mathcal{N}(\vec{x}_i|\vec{y}, \mtx{W} + \mtx{\Sigma}_i)
    \end{aligned}
    \label{eq:plda-up}
\end{equation}
By setting $\mtx{\Sigma}$ to zero the conventional PLDA model is recovered.

In practice, we use a PLDA model with \emph{spherical covariances}, \emph{i.e.} with $ \mtx{B} = \sigma_\text{B}^2\mtx{I}$, $\mtx{W} = \sigma_\text{W}^2\mtx{I}$, and $ \mtx{\Sigma}_i = \sigma_i^2\mtx{I}$, where $\mtx{I}$ denotes an identity matrix. Here, $\sigma_i^2$ is the inverse of precision computed from the embedding magnitude by \eqref{eq:norm-transform}.

\setcounter{footnote}{0} 
We modified the publicly available implementation of \cite{landini2020bayesian} to support uncertainty propagation and used the original implementation as a baseline\footnote{\scriptsize \url{https://github.com/BUTSpeechFIT/VBx}}, further referred to as VBx.




\subsection{Experiment Setup}

We conducted the experiments using the popular AMI \cite{ami_corpus} and VoxConverse \cite{chung20_voxconverse} corpora. 
We followed the development/evaluation split for the AMI corpus from the official website \footnote{\scriptsize \url{https://groups.inf.ed.ac.uk/ami/corpus/datasets.shtml}} and for the VoxConverse we used dev v0.0.1/eval v0.0.2 sets from the official GitHub repository\footnote{\scriptsize \url{https://github.com/joonson/voxconverse}}, where the development set is used only for tuning the parameters of the clustering algorithm. That is, we used the same embedding extractors as described in the previous section without fine-tuning on the AMI or VoxConverse data.



We used a simple speaker diarization pipeline including the following steps: voice activity detection (VAD), overlapped speech detection (OSD; both from \cite{Bredin2020, Bredin2021}), fixed-length segmentation, clustering, and post-processing. 
The post-processing includes merging the adjacent sub-segments from the same speaker and distributing the overlapped segments equally among the adjacent segments with different speakers.
We extracted embeddings from segments of length $1.5$ sec with $0.75$ overlap within the boundaries computed by VAD. 
These embeddings were further clustered using one of the following algorithms: 

\textbf{AHC}: agglomerative hierarchical clustering (AHC) algorithm described in \cite{landini2020bayesian}. The stopping threshold was tuned on the development set. 

\textbf{VBx}: clustering model based on Bayesian HMM from \cite{landini2020bayesian} with an initial assignment by the AHC output. Similar to AHC, the VBx parameters were tuned on development set.

\textbf{VBx-UP}: the proposed modification of VBx with uncertainty propagation that uses reciprocals of magnitudes as segment-specific variances as shown in \eqref{eq:plda-up}. %

Finally, we reassign labels for segments with overlaps detected by the OSD module. We used a simple heuristic: we assign two closest speakers by time to every segment marked by OSD.

\subsection{Evaluation results}
We report the results using the \emph{diarization error rate} (DER) \cite{DER_explained} and \emph{Jaccard error rate} (JER) \cite{DIHARD-2} metrics. In this experiment, we did not use the forgiveness collar. In addition, the speaker overlap regions were included during scoring at the evaluation step. We followed this evaluation setup in all the reported experiments except the one shown in Table \ref{diar:but_like_table}.
 
First, each of the three variants of the proposed two-step method outperformed the baseline in the DER and JER metrics on the AMI Headset-Mix channel. In the evaluation results provided in Table \ref{tab:two_step_comparison}, we found out that the Algorithm 2.1 provides better results than the other two. Therefore, we report the results only for the Algorithm 2.1 in Tables \ref{diar:but_like_table}, \ref{diar:embedders_comparison} and \ref{diar:clustering_comparison}. 

Here and further adding a postfix to the clustering algorithm name indicates that the two-step clustering procedure was used. The specific variant is shown in brackets with the Algorithm 2.1 being a default option.
 
\setlength{\tabcolsep}{3.0pt}   

\begin{table}[!ht]
\centering
\begin{tabular}{lcc}
\thickhline
\multicolumn{1}{c}{\textbf{Clustering}} & \textbf{DER, \%} & \textbf{JER, \%} \\ \thickhline
AHC                                             & 22.75            & 28.29            \\ \thickhline
AHC-2step                                         & \textbf{13.13}   & \textbf{23.62}   \\ \hline
AHC-2step (2.2)                                         & 19.07            & 25.75            \\ \hline
AHC-2step (2.3)                                         & 17.04            & 26.68            \\ \thickhline
\end{tabular}
\caption{Comparison of two-step clustering algorithms based on ResNet-MagFace embedding extractor and AHC clustering algorithm. The evaluation was performed on the AMI Headset-Mix channel without any forgiveness collar and with overlaps included.}
\label{tab:two_step_comparison}
\end{table}

\begin{table}[!ht]
\centering
\footnotesize
\begin{tabular}{c|l|c|cc|c|c}
\thickhline
\textbf{Dataset}                       & \textbf{Clustering}        & \textbf{OSD}         & \multicolumn{2}{c|}{\textbf{Setup}}                     & \textbf{DER, \%} & \textbf{JER, \%}       \\ \hline
\textbf{}                              &                            &                      & \multicolumn{1}{c|}{\textbf{collar}} & \textbf{overlap} &                  &                        \\ \thickhline
\multirow{12}{*}{\textbf{\rotatebox[origin=c]{90}{AMI}}}         & \multirow{3}{*}{AHC}       & \multirow{3}{*}{No}  & 0.25                                 & No               & 6.89             & \multirow{3}{*}{32.59} \\
                                       &                            &                      & 0.25                                 & Yes              & 16.58            &                        \\
                                       &                            &                      & 0                                    & Yes              & 27.27            &                        \\ \cline{2-7} 
                                       & \multirow{3}{*}{AHC-2step} & \multirow{3}{*}{No}  & 0.25                                 & No               & 2.76             & \multirow{3}{*}{28.12} \\
                                       &                            &                      & 0.25                                 & Yes              & 11.74            &                        \\
                                       &                            &                      & 0                                    & Yes              & 20.66            &                        \\ \cline{2-7} 
                                       & \multirow{3}{*}{AHC}       & \multirow{3}{*}{Yes} & 0.25                                 & No               & 6.89             & \multirow{3}{*}{28.29} \\
                                       &                            &                      & 0.25                                 & Yes              & 13.91            &                        \\
                                       &                            &                      & 0                                    & Yes              & 22.75            &                        \\ \cline{2-7} 
                                       & \multirow{3}{*}{AHC-2step} & \multirow{3}{*}{Yes} & 0.25                                 & No               & 2.76             & \multirow{3}{*}{23.62} \\
                                       &                            &                      & 0.25                                 & Yes              & 7.07             &                        \\
                                       &                            &                      & 0                                    & Yes              & 13.13            &                        \\ \thickhline
\multirow{12}{*}{\textbf{\rotatebox[origin=c]{90}{VoxConverse}}} & \multirow{3}{*}{AHC}       & \multirow{3}{*}{No}  & 0.25                                 & No               & 5.96             & \multirow{3}{*}{28.01} \\
                                       &                            &                      & 0.25                                 & Yes              & 7.61             &                        \\
                                       &                            &                      & 0                                    & Yes              & 11.83            &                        \\ \cline{2-7} 
                                       & \multirow{3}{*}{AHC-2step} & \multirow{3}{*}{No}  & 0.25                                 & No               & 4.24             & \multirow{3}{*}{28.53} \\
                                       &                            &                      & 0.25                                 & Yes              & 5.84             &                        \\
                                       &                            &                      & 0                                    & Yes              & 9.41             &                        \\ \cline{2-7} 
                                       & \multirow{3}{*}{AHC}       & \multirow{3}{*}{Yes} & 0.25                                 & No               & 5.96             & \multirow{3}{*}{25.07} \\
                                       &                            &                      & 0.25                                 & Yes              & 6.74             &                        \\
                                       &                            &                      & 0                                    & Yes              & 11.39            &                        \\ \cline{2-7} 
                                       & \multirow{3}{*}{AHC-2step} & \multirow{3}{*}{Yes} & 0.25                                 & No               & 4.24             & \multirow{3}{*}{24.79} \\
                                       &                            &                      & 0.25                                 & Yes              & 5.23             &                        \\
                                       &                            &                      & 0                                    & Yes              & 9.21             &                        \\ \thickhline
\end{tabular}
\caption{Comparison of speaker diarization methods based on different clustering algorithms with and without OSD module. The evaluation was performed on the AMI test and VoxConverse test v0.0.2.}
\label{diar:but_like_table}
\end{table}

\begin{table}[!ht]
\footnotesize
\centering
\begin{tabular}{c|c|l|c|c}
\thickhline
\textbf{Dataset}                      & \textbf{Model}               & \textbf{Clustering} & \textbf{DER, \%} & \textbf{JER, \%} \\ \thickhline
\multirow{6}{*}{\textbf{\rotatebox[origin=c]{90}{AMI}}}         & \multirow{2}{*}{ResNet-ArcFace}     & AHC                 & 21.68            & 25.28            \\
                                      &                              & AHC-2step             & 13.06            & 24.11            \\ \cline{2-5} 
                                      & \multirow{2}{*}{ResNet-MagFace}     & AHC                 & 22.75            & 28.29            \\
                                      &                              & AHC-2step             & 13.13            & 23.62            \\ \cline{2-5} 
                                      & \multirow{2}{*}{Large ResNet-MagFace} & AHC                 & 20.15            & 25.95            \\
                                      &                              & AHC-2step             & 15.87            & 24.34            \\ \thickhline
\multirow{6}{*}{\textbf{\rotatebox[origin=c]{90}{VoxConverse}}} & \multirow{2}{*}{ResNet-ArcFace}     & AHC                 & 10.71            & 24.40             \\
                                      &                              & AHC-2step             & 7.95             & 25.28            \\ \cline{2-5} 
                                      & \multirow{2}{*}{ResNet-MagFace}     & AHC                 & 11.39            & 25.07            \\
                                      &                              & AHC-2step             & 9.21             & 24.79            \\ \cline{2-5} 
                                      & \multirow{2}{*}{Large ResNet-MagFace} & AHC                 & 8.87             & 20.47            \\
                                      &                              & AHC-2step             & 7.68             & 22.65            \\ \thickhline
\end{tabular}
\caption{Comparison of speaker diarization methods based on different embedder models with Oracle VAD and OSD modules. The evaluation was performed on the AMI Headset-Mix channel and the VoxConverse test v0.0.2.}
\label{diar:embedders_comparison}
\end{table}

\begin{table}[]
\centering
\begin{tabular}{c|l|c|c}
\thickhline
\textbf{Dataset}             & \textbf{Clustering} & \textbf{DER, \%}   & \textbf{JER, \%}   \\ \thickhline
\multirow{6}{*}{\textbf{\rotatebox[origin=c]{90}{AMI}}}         & AHC                      & 22.75          & 28.29          \\
                             & AHC-2step                  & \textbf{13.13}          & \textbf{23.62} \\ \cline{2-4} 
                             & VBx                      &  16.82          & 23.33          \\
                             & VBx-2step                  & \textbf{11.15}          & \textbf{22.38}          \\ \cline{2-4} 
                             & VBx-UP                   & 17.60          & 25.50          \\
                             & VBx-UP-2step               & \textbf{10.87} & \textbf{22.25}          \\ \thickhline
\multirow{6}{*}{\textbf{\rotatebox[origin=c]{90}{VoxConverse}}} & AHC                      & 11.39          & 25.07          \\
                             & AHC-2step                  & \textbf{9.21}           & \textbf{24.79}          \\ \cline{2-4} 
                             & VBx                      & 10.73           & 24.33           \\
                             & VBx-2step                  & \textbf{9.76}           & \textbf{23.93}           \\ \cline{2-4} 
                             & VBx-UP                   & 8.28           & 26.93           \\
                             & VBx-UP-2step               & \textbf{6.99}           & \textbf{23.14}           \\ \thickhline
\end{tabular}
\caption{Comparison of speaker diarization methods based on different clustering algorithms with Oracle VAD and OSD modules. The evaluation was performed on the AMI Headset-Mix channel and VoxConverse datasets. The embedding extractor is the ResNet-MagFace.}
\label{diar:clustering_comparison}
\end{table}

Table \ref{diar:but_like_table} provides the results of AHC and AHC-2step methods with three evaluation setups with different settings for the forgiveness collar and overlap parameters. The proposed two-step approach allowed to achieve a $60\%$, $49\%$ and $42\%$ relative reduction in DER with different evaluation setups (from the most forgiving to the most strict, respectively) on the AMI corpus. The JER metric was improved by $17\%$ and $13\%$ in cases with and without the OSD module, correspondingly.
On the VoxConverse data the DER metric has similar behavior ($17\%$ relative reduction), however, surprisingly, the JER metric remained stationary. Additionally, the two-step approach with the overlapped speech detection module achieved higher relative improvement ($49\%$) than the setup without the overlapped speech detection ($24\%$).

Next, Table \ref{diar:embedders_comparison} shows a comparison of different embedding extractors. The ResNet-Arcface model performed slightly better than the ResNet-MagFace, but the relative improvement in both DER and JER of the AHC-2step algorithm is higher for the ResNet-MagFace on the AMI data ($40\%$ vs $42\%$ and $4\%$ vs $17\%$ in DER and JER, respectively). Also, the Large ResNet-MagFace performed the best among all setups with the one-step AHC algorithm, but it achieved the least relative improvement in the DER metric. A similar observation can be made for the VoxConverse data.

Finally, Table \ref{diar:clustering_comparison} demonstrates experimental results for different clustering algorithms. One can observe that the two-step procedure achieved lower DER and JER for all the reported clustering algorithms. Also, uncertainty propagation with the two-step clustering further helped to improve DER and JER metrics compared to the original VBx algorithm especially in combination with the two-step clustering.



\section{Analysis}
\label{sec:analysis}
In this section, we examine the quality propagation abilities of the learned magnitudes for the ResNet-MagFace model.

\begin{figure}[!h]
 \centering
 \includegraphics[trim=0mm 6mm 0mm 4mm, clip, scale=0.33]{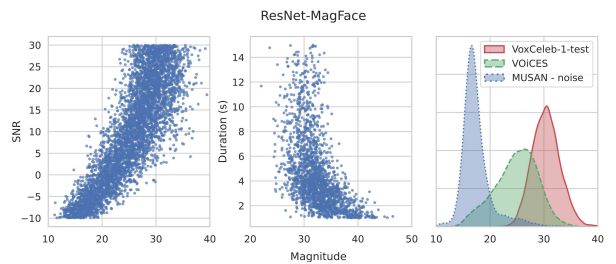}
 \caption{Left: SNR versus magnitude (VC1-test, additive white noise). Center: duration versus magnitude (VC2-test). Right: distributions of the embedding magnitudes for three different datasets. All the results are reported for the ResNet-MagFace model. Best viewed in color.}
 \label{fig:qualitative}
\end{figure}

First, we augmented the recordings from the VoxCeleb1 test with additive white Gaussian noise with random SNR selected uniformly in $[-10;30]$. Figure \ref{fig:qualitative} (Left) demonstrates a scatter plot of SNRs versus magnitudes for the embeddings extracted from the noisy recordings. This result confirms the interpretation of the embedding magnitudes as quality scores.

Further, we examined the relation of the utterance duration to the embedding magnitude. To reduce the effect of other possible sources of uncertainty, we used the relatively clean VoxCeleb data (VC2-test). We picked a set of 2000 recordings and cropped them at a random moment selected uniformly in $[1\text{s};15\text{s}]$. Figure \ref{fig:qualitative} (Center) depicts the results. Surprisingly, the magnitudes become larger for shorter segments, which somewhat contradicts the interpretation of magnitudes as confidence scores. However, this effect becomes more apparent only for small durations and can be easily compensated, \emph{e.g.}, by introducing the duration-dependent adjustment, which we describe in section \ref{subsec:norm2prec}.

Finally, we estimated the distribution of magnitudes for three different sets of recordings: the noise part of MUSAN corpus \cite{musan2015}, the VoxCeleb1 test and the VOiCES Eval set. From Figure \ref{fig:qualitative} (Right) one can see that the learned magnitudes allow for more accurate separation between the clean speech (VoxCeleb) and non-speech (MUSAN) than those from the extractors shown on Figure \ref{fig:clova-ecapa}. Therefore, magnitudes can also be utilized for the out-of-distribution detection in the embedding space. 
This property can possibly be employed as an additional layer of defense to compensate for failures of a voice activity detector, as suggested by \cite{kwon2021look}. 

\begin{figure}[htb] 
 \centering
 \includegraphics[trim=0mm 3mm 0mm 2mm, clip, scale=0.5]{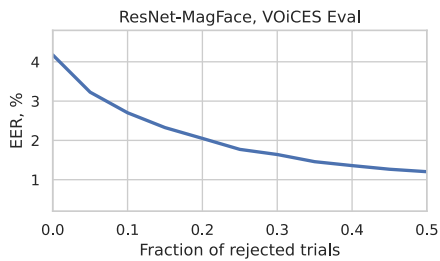}
 \caption{Results of speaker verification with a reject option for the VOiCES Eval set. The plot presents EER vs. the fraction of rejected trials. Trials with the lowest sum of embeddings magnitudes are discarded.}
 \label{fig:rc}
\end{figure}

We also conducted a quantitative evaluation of the ability of magnitudes to serve as quality measures. We considered the speaker verification task, where a recognizer can reject a trial if there is a high risk of making a wrong prediction. In detail, given a trial list, we computed the confidence scores as a sum of magnitudes of embeddings forming a trial pair. Then, we discarded a subset of trials with the lowest confidence scores and computed the performance metrics using the remaining trials. For this scenario we adopt the VOiCES Eval set, which provides the recordings with varying quality. Fig. \ref{fig:rc} illustrates how the equal error rate (EER) depends on the fraction of rejected trials.

The magnitudes emerge useful for filtering out trials with incorrect predictions. Even with the rejection fraction of 0.2, the EER on the VOiCES Eval test is reduced by half, revealing the high capability of magnitudes for quality assessment.

\section{Conclusion}
We presented a speaker embedding extractor providing interpretable magnitudes suitable for out-of-the-box quality assessment or propagating the quality information into the back-end model.
We proposed a new probabilistic embedding representation that can be obtained from a pre-trained extractor network with almost no additional effort.
We evaluated the proposed model for the speaker verification task and achieved up to 20\% relative improvements over the magnitude-agnostic baselines, for both clean and noisy test sets. In addition, we proposed a two-step clustering algorithm for speaker diarization, which improved the DER and JER metrics relatively up to $60\%$ and $17\%$, respectively, on two popular public benchmarks. Moreover, the proposed modification of the VBx model with uncertainty propagation achieved better performance compared to the original model. It is worth noting that the two-step approach is faster than the one-step as it takes only a subset of embeddings to perform computationally expensive clustering. We also provided a comprehensive analysis of magnitudes properties, which highlights the potential application for speech quality assessment tasks. We plan to make our models and evaluation protocols available online.



In the future work, we aim at training a magnitude-aware embedding extractor from scratch to get rid of the ad hoc duration variability compensation transform.
Another direction includes integrating the magnitude-based quality assessment into the two-step pipelines based on the target-speaker VAD such as \cite{He2021-tsvad}. 


\bibliographystyle{IEEEbib}
\bibliography{Odyssey2022_BibEntries}

\end{document}